\documentclass[sigconf]{acmart}

\AtBeginDocument{%
  \providecommand\BibTeX{{%
    \normalfont B\kern-0.5em{\scshape i\kern-0.25em b}\kern-0.8em\TeX}}}

\copyrightyear{2024}
\acmYear{2024}
\setcopyright{rightsretained}
\acmConference[C\&C '24]{Creativity and Cognition}{June 23--26, 2024}{Chicago,
IL, USA}
\acmBooktitle{Creativity and Cognition (C\&C '24), June 23--26, 2024, Chicago,
IL, USA}
\acmDOI{10.1145/3635636.3656187}
\acmISBN{979-8-4007-0485-7/24/06}

\usepackage{subfig}
\usepackage[inkscapelatex=false]{svg}
\usepackage{csquotes}
\usepackage{caption}
\usepackage{tabularx}
\usepackage{array}

\begin{document}
 \newenvironment{myquote}{\list{}{\leftmargin=0.02\textwidth \rightmargin=0.02\textwidth}\item[]}{\endlist}
        \newcommand*{\participant}[1]{{\textit{\small{\fontfamily{cmss}\selectfont{(#1)}}}}}
        \newcommand*{\quoted}[1]{{\small{\fontfamily{cmss}\selectfont{#1}}}}
        \newcommand{\squote}[2]{\begin{myquote}\quoted{#2 \participant{#1}}\end{myquote}}
        \newcommand*{\quotedtext}[1]{\begin{myquote}\small{#1}\end{myquote}}

    \newcommand{\centered}[1]{\begin{tabular}{l} #1 \end{tabular}}

    \newcolumntype{L}[1]{>{\raggedright\let\newline\\\arraybackslash\hspace{0pt}}m{#1}}


\title{\textit{Ai.llude}: Encouraging Rewriting AI-Generated Text to Support Creative Expression}


\author{David Zhou}
\affiliation{%
  \institution{University of Illinois Urbana-Champaign}
  \streetaddress{201 N Goodwin Ave}
  \city{Urbana}
  \country{USA}}
\email{david23@illinois.edu}

\author{Sarah Sterman}
\affiliation{%
  \institution{University of Illinois Urbana-Champaign}
  \streetaddress{201 N Goodwin Ave}
  \city{Urbana}
  \country{USA}}
\email{ssterman@illinois.edu}

\renewcommand{\shortauthors}{David Zhou and Sarah Sterman}

\begin{abstract}

In each step of the creative writing process, writers must grapple with their creative goals and individual perspectives. This process affects the writer’s sense of authenticity and their engagement with the written output.
Fluent text generation by AIs risks undermining the reflective loop of rewriting.
We hypothesize that deliberately generating imperfect \textit{intermediate text} can encourage rewriting and prompt higher level decision making.  
Using logs from 27 writing sessions using a text generation AI, we characterize how writers adapt and rewrite AI suggestions, and show that intermediate suggestions significantly motivate and increase rewriting. 
We discuss the implications of this finding, and future steps for investigating how to leverage intermediate text in AI writing support tools to support ownership over creative expression. 

\end{abstract}

\begin{teaserfigure}
    \includegraphics[width=\textwidth]{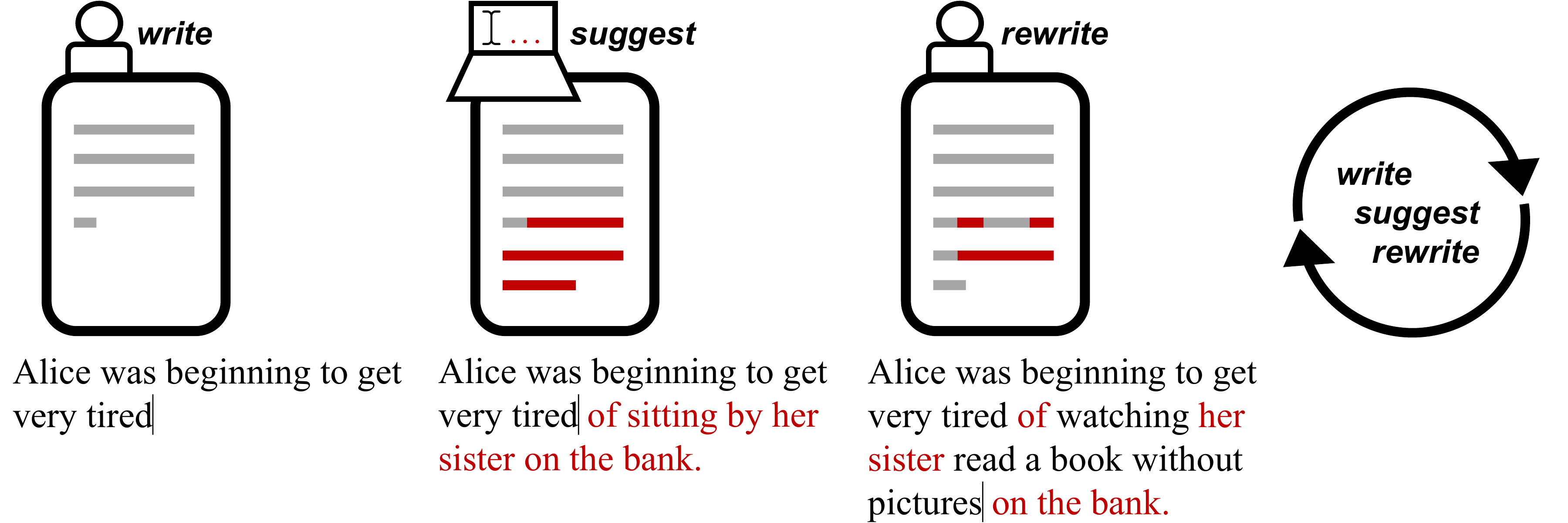}
    \caption{
        Rewriting is an important stage of creative writing, influencing self-expression, creative goals, and final output. This paper studies how writers adapt AI suggestions and evaluates the effect of \textit{rewriting AI-generated suggestions} on writing process. Figure text: adapted from Alice in Wonderland (Lewis Carroll).
    }
    \label{fig:teaser}
    \Description{Figure 1 depicts the flow of requesting and integrating AI-generated suggestions in creative writing. At the left of the image is a depiction of a rectangle with horizontal grey lines inside it to indicate a page of written text, and icon of a person and the word "write" on top. The text (in black) "Alice was beginning to get very tired," followed by a line representing a text caret, is below the page. An arrow connects this depiction to a second icon of a page, now ending with horizontal red lines, indicating the presence of suggested AI text, and an icon of a laptop with the word "suggest" on top. The text (in black) "Alice was beginning to get very tired," followed by a line representing a text caret, then the text (in red) "of sitting by her sister on the bank.," followed by a line representing a text caret, is below the page. From this icon, another arrow leads to the third icon of a page, now with grey lines breaking up the red lines, indicating the user has rewritten part of the AI text. The text (in black) "Alice was beginning to get very tired" (in red) "of" (in black) "watching" (in red) "her sister" (in black) "read a book without pictures," followed by a line representing a text caret, then the text (in red) "on the bank." There is an icon of a person with the word "write suggest rewrite" on top. A final arrow connects this depiction to the fourth icon of a page, now containing only grey lines. There are two arrows in a circle representing a cycle beneath the final figure to represent that the process repeats.}
\end{teaserfigure}



\maketitle

\section{Introduction}
Creative writing is a form of self-dialogue \cite{hunt_self_1998, cramer_2007_I}.
Through the process of writing, writers reflect on not only language, but also aesthetics, imagination, and identity.
Advances in language models have led to waves of new intelligent writing assistants designed to make writing easier by generating fluent text \cite{sudowrite, hemingwayapp}. While these tools can speed up the generation of content or help writers overcome writer's block ~\cite{singhstolenelephant22, clarkslogansstories18, skjuve2023user}, it is important to consider how computational tools, particularly ones that take over self-expression, affect personal writing processes and goals. 

One significant concern is that by removing the writer from the process of making difficult choices about ideas and words, such tools may also reduce the space for deep thinking that occurs through struggle.
Feelings of authenticity and ownership are associated with the struggle to create inventive ideas \cite{Belk88, Murray91} and the investment of effort, attention, and time into the textual artifact \cite{csikszentmihalyi_meaning_1981}. 
In recent work, the need for user control and ownership have emerged as potential design guidelines for intelligent writing tools \cite{biermann22, Gero19metaphoria}.

Therefore a key question arises for the design of intelligent writing tools: what aspects of intelligent writing assistance increase or undermine writers' control over the work, and how do we balance the benefits of intelligent writing assistants, such as improvements in writing efficiency or democratization of writing skills, with the need for creative writers to have space for critical reflection?
We hypothesize that rewriting can be an important dimension for understanding how AI text generators influence feelings of control. As a first step, this paper looks at rewriting AI-generated text as a way to support control over writing process and output.

Rewriting is an essential part of the creative writing process; during rewriting, language is refined, removed, or inserted in order to change, clarify, or expand the ideas of the text. In doing so, the writer grapples both with ideas and the effective expression of those ideas through words and structure. 
Prior work has studied the relationship between rewriting AI-generated text and psychological ownership in limited forms \cite{leecoauthor22}. 
However, significant questions remain around the effect of rewriting on writer's process, understanding why and how writers rewrite AI-generated text, and defining rewriting in the context of AI generation.
In this work, we address the following questions: 
\textit{How can AI suggestions be designed to encourage the rewriting of AI-generated text?} and \textit{How does the rewriting of AI-generated text affect the writer's personal writing process?}

To explore the rewriting of AI-generated text, this paper presents a study of 27 creative writers using a custom text editor that provides two types of AI-generated suggestions: first, a \textit{fluent continuation} to follow the user's text; second, an \textit{intermediate suggestion} consisting of multiple fragmentary ideas.
For the first, we request a completion of the user's text without additional prompts; the second is created by concatenating four potential ideas about plot and setting into a single continuation. 
Fluent continuations are designed to be grammatical and coherent to context.
Intermediate suggestions are meant to provide relevant content that cannot be directly incorporated into the text.
Instead, suggestions must deliberately be altered to become grammatical and coherent. 
We hypothesize that shifting the focus of generative writing assistants away from creating perfect output to creating intermediate text that encourages rewriting will support the personal and reflective aspects of creative writing.

We use three metrics to capture different aspects of rewriting behaviors: \textit{remaining AI text}, \textit{sentence embedding similarity}, and \textit{number of user edits}. We use these metrics, along with qualitative analysis, to evaluate rewriting behaviors and writing engagement, as well as characterizing the difference between these two completion paradigms.  
We find that writers tend to keep less of intermediate suggestions compared to fluent continuations. Despite being perceived as more flawed, we find that writers are able to adopt intermediate suggestions more diversely; rather than simply for writing continuation, writers use these suggestions to help brainstorm ideas, even if none of the original suggestion remains in the final story. 

This paper contributes:
\begin{enumerate}
    \item The concept of \textit{intermediate text} and an implementation of an \textit{intermediate suggestion}, a paradigm for engaging creative writers in editing AI-generated text.
    \item Results from a 27-person study of writing with \textit{intermediate suggestions}:
    \begin{itemize}
        \item characterizing rewriting behaviors on AI-generated text in creative writing
        \item showing intermediate suggestions increase rewriting
        \item showing the effects of suggestion design on writing process: fluent continuations support writing ease, while intermediate suggestions support ideation and reflection.
    \end{itemize}
    \item Discussion of designing generative writing assistants that center writers' process and ownership.
\end{enumerate}
\section{Related Work}
\subsection{Creative Writing and Rewriting}
Creative writing is deeply personal, open-ended, and connected to self-expression, making the process of how we write an important subject of study.
There are no unambiguous definitions of acceptable outcomes for creative tasks, nor are there obvious and predetermined paths to creative goals \cite{amabilesocialpsy83}. Instead, the process of seeking a creative goal can be valuable in itself and essential to the character of the final output. Writing is an act of \textit{making} meaning, not finding it \cite{flowerhayesrhetoricalproblem80}, and the act of writing can reciprocally shape the author as well \cite{Murray91}. 
When we consider that the tools and environments we use shape our process, not only in how we take actions but even in what actions we can imagine or are available to take \cite{dalsgaard2017instruments, flowerhayes81}, it becomes essential to ask how our writing support tools affect writers' process of expression. 

Rewriting is a vital step in the writing process.
Rewriting is not only the act of writing again \cite{rewritingOED}; it is intertwined with the overall effort to \textit{make it right} — to move from word to phrase to sentence to paragraph then back to words, to see the need for variety and balance, firmer structure, and more appropriate form
\cite{murray83themakerseye}.
Writing, as a process of meaning-making and discovery, can be characterized as a reactive process between  writer and textual artifact \cite{flowerhayes81, galbraith04revision}. 
This process involves reading and rereading, as well as writing and rewriting \cite{murray83themakerseye}, a form of hands-on thinking \cite{lucy80childrens}. In his book \textit{Writing Without Teachers}, Elbow describes writing as words gradually changing and evolving \cite {elbow1998writing}. Authors of classic literature Leo Tolstoy and Roald Dahl describe rewriting as a reflective practice, one where writing goals develop and completely change over time \cite{murray83themakerseye}; Ernest Hemingway rewrote the ending to his first bestseller \textit{Farewell to Arms} 
thirty-nine times \cite{hemingway39}. Since rewriting is a reciprocal process, it is important to understand what factors influence rewriting and what factors motivate it in the author.

Most writing now is done on computers with some form of word processor. Since tools fundamentally shape process, we must consider how they affect rewriting, and therefore meaning-making and self-expression.
The effects of word processors on writing process have been studied for decades; for example, writing with a word processor has led to more local revisions (word, phrase, sentence level) compared to writing by hand \cite{collier83wordprocessor, dave10conceptsdrafting}; for novice writers, these changes tended to be lexical \cite{Sommers80RevisionSO} while experienced writers tended to make global changes \cite{hill91revising}. The effects of AI text generation tools are still uncertain, especially as the capability of AI systems to output fluent text has recently increased dramatically. 
In this paper, we explore how two variants of text generator outputs affect rewriting behaviors and how writers perceive AI writing in context of their personal writing processes.

\subsection{AI Writing Support Tools}
\label{sec:writing-support-tools}

Writing support tools have been a mainstay of creativity support research and development in HCI since the early days of computing \cite{peterson80spellcheck}. 
Recent works have proposed specialized tools to support greater control over expression, such as word finding and writing style \cite{gero19stylisticthesaurus, Gabrielinkwellstyle15}. Studies have shown that writers have diverse expectations when integrating computational assistance into their existing creative writing process \cite{booten21poetry, Ippolito22}. 
Recent studies have also explored the need for self-reflection, such as by watching real-time writing replays \cite{carrera22watch}, and conversing with 
a character development tool to explore deeper characterization \cite{schmitt21characterchat}.

In recent years, a rapid increase in fluency and capability of large language models has driven a surge of AI-supported writing tools. Here we discuss three types of outputs for generative writing tool design: \textit{fluent multipurpose generation}; \textit{task-specific targeted suggestions}; and \textit{resources for interpretation}. In the next section, we discuss the design goal of encouraging \textit{rewriting} in relationship to AI features.   

\textit{Fluent multipurpose generation:}
Large language models now have the capability to generate contextually relevant and fluent text.  One genre of AI writing support tools use this ability to generate completions or suggestions that are inserted directly into the user's text \cite{leecoauthor22, clarkslogansstories18, sudowrite, roemmele_creative_2015, roemmele_automated_2018, yuan_wordcraft_2022}. These suggestions are designed to be relevant, grammatically correct and potentially included into the final output without modification. Such tools offer completions during the writing process, and are motivated by goals like helping the user overcome writer's block, provide new ideas, or speed up the writing process~\cite{singhstolenelephant22, clarkslogansstories18, skjuve2023user}.

\textit{Task-specific targeted suggestions:}
A second genre of AI writing support tools focuses on supporting one particular type of writing task, providing targeted suggestions rather than general text completion. For example,
Metaphoria \cite{Gero19metaphoria} helps writers create metaphors by suggesting connections to ideas and words. 
Inkwell provides stylistic variations of texts \cite{Gabrielinkwellstyle15} to assist style goals in poetry. A proposed character backstory generation system \cite{clarkcharacterbackstory22} helps game designers create convincing characters in immersive environments. By targeting specific writing tasks, these tools intervene only in a limited part of the writing process. 

\textit{Resources for interpretation:}
\label{sec:resources-for-interpretation}
Writing support tools can also provide resources for interpretation as their text generation goal. Rather than providing text meant to be included in the output, these tools provide annotations, feedback, or inspirational media. For example, multimodal generation provides writers non-text suggestions that force the writer to evaluate and interpret provided feedback 
\cite{singhstolenelephant22}. Hai et al. propose providing writers with paragraph summaries as margin annotations to support revision \cite{haibeyondtextgeneration22}. These AI generations are designed to motivate writing processes while leaving greater control and responsibility over choices about the text itself to the user. 

In each of these tools, the usefulness and perceived quality of its suggestions are often limited by the technical capabilities of the underlying language model. Despite these shortcomings, AI text generators do not necessarily require ideal coherency to support creativity processes. Literary cut-up techniques, a method of creating new texts by combining the fragments of existing texts, employ chance to generate novelty \cite{robinson2011shift, hart17autonomous}. The stochasticity of language models may also function as a source for inspiration — to inspire writers to make cognitive leaps between elements of semantic relevance in order to further their creative goals \cite{singhstolenelephant22}.

\subsection{Motivating Rewriting}

Flower and Hayes identify the task environment as an influential external factor of the cognitive writing process \cite{flowerhayes81}. AI writing support features alter the task environment and affect both engagement with writing and expression of personal perspective. 
Generative writing assistants that directly insert text into the writing change the task environment and have the potential to negatively affect feelings of engagement, particularly if the provided suggestions are very good \cite{Gero19metaphoria}.  
Jakesch et al. show that suggestions provided by an opinionated language model alter the opinions writers express in their work \cite{jakesch2023cowriting}.
To design AI writing assistants that can help with writing tasks without undermining the writer's exploration of meaning and expression, we must consider how to increase engagement and provide greater control to the writer over their process.  This paper takes up \textit{rewriting} as one design direction to support ownership and expression.  

Rewriting is an integral piece of the writing process, an opportunity to find new and surprising ideas and learn from one's own expression \cite{morley2007cambridge}.
In reflective practice, the practitioner changes their understanding of the problem space and updates the actions they take through the process of doing \cite{schon1983reflective}; for creative writers, rewriting can be a powerful source of reflection-in-action.  The connection between rewriting and the construction of goals and self-expression underlies our decision to investigate rewriting as a potential mediator of psychological ownership. 

Lee et al. performed a preliminary analysis of rewriting and ownership by looking at rewriting behaviors across the entire text \cite{leecoauthor22}. They found a weak, insignificant relation between the number of edits, approximated by the number of selection and delete events, across the entire text and ownership. However, they were not able to investigate the rewriting of AI suggestions specifically. We seek to deepen our understanding of the relationship between rewriting and ownership by evaluating rewriting specifically done on AI-generated text with a more extensive survey assessing control over self-expression.

We also seek to understand how to create AI-generated text that encourages rewriting. We build on the motivation of systems that generate resources for interpretation (Section \ref{sec:resources-for-interpretation}): providing material for the user to interpret, rather than solutions to incorporate. We design a system to generate intermediate suggestions, which must be rewritten by the user before it can be coherently incorporated into the story. While older AI tools tend to focus on the benefit of serendipity to the writing process and are limited by the coherence of the models  \cite{Gabrielinkwellstyle15, roemmele_automated_2018, calderwood20hownovelists}, the recent increase in the capability of language models to generate fluent and relevant text enables us to design deliberate alterations to coherency.

\begin{figure*}%
    \centering
    \includegraphics[width=0.95\textwidth]{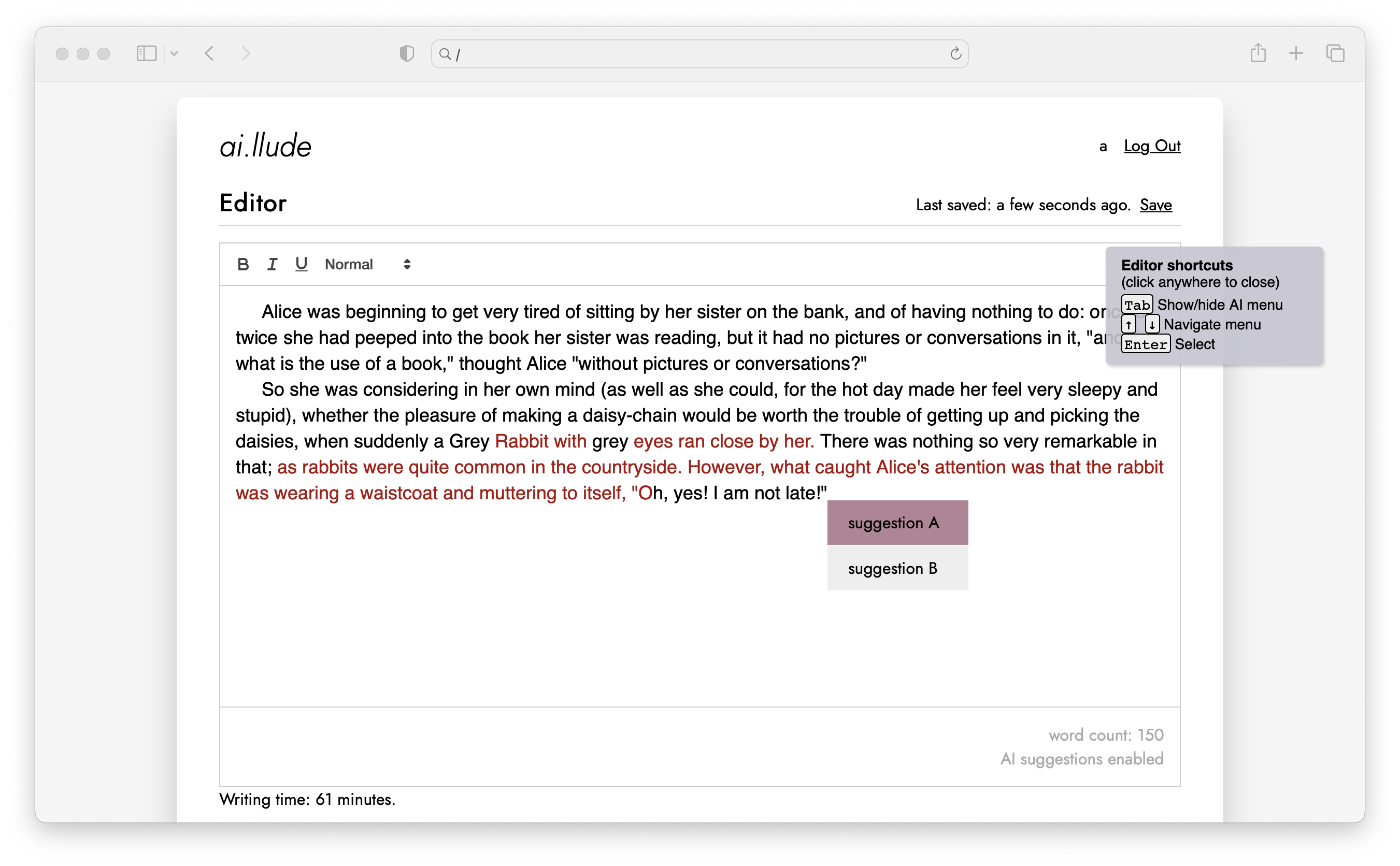} 
    \caption{Interface of \textit{ai.llude}. The AI menu opens when the tab key is pressed and allows writers to request suggestions. User text is displayed in black and AI generated text in purple. The word count and writing time are displayed at the bottom of the editor. The editor provides basic styling controls such as bold, italic, and underline. Figure text: Alice in Wonderland (Lewis Carroll).}%
    \label{fig:interface}%
    \Description{Figure 2 is a screenshot of the user interface of ai.llude, a custom web-based text editor. It contains a large text entry box where writers draft stories, containing an excerpt of Alice in Wonderland (Lewis Carroll). The AI menu, which contains two options labeled Suggestion A and Suggestion B, is displayed at the end of the text. It contains a mix of black text and purple text. The Editor shortcuts menu, a small floating window, is shown. Users can press tab to show/hide the AI menu, use the arrow keys to navigate the menu, and press enter to request the suggestion. The elapsed time of the writing session, word count, and text showing that AI suggestions are enabled are placed at the bottom. On the top of the editing window is a style strip for stylizing text. The interface shows when the writing was last saved, a save button, and a log out button to quit the session.}
\end{figure*}

\section{Design of ai.llude}

To study rewriting behaviors, we built a system for data collection to explore how we might encourage rewriting. \textit{Ai.llude} is an instrumented text editor with AI support designed to encourage rewriting. It is designed to simulate a realistic writing environment and provides a streamlined interface for working with AI suggestions, as well as extensive logging capabilities.

\subsection{Design Goals}
Past work has identified
\textbf{interactivity} and \textbf{control} in AI writing assistants as design characteristics that affect perceptions of engagement with the assistant and \textbf{ownership} over the writing artifact \cite{Gero19metaphoria, biermann22}. These principles are not exclusive to digital writing assistants. For human-to-human design collaboration, defining scope and encouraging conversation make creation and collaboration more clear and engaging \cite{gibbons_design_nodate}, while the emphasis of voice and identity in writing contexts in general provides the writer with greater agency and expression \cite{bishop_colors_1994}. 

To address the need for greater {interactivity} and {control}, we explore the paradigm of generating intermediate text. In contrast to \cite{singhstolenelephant22, haibeyondtextgeneration22}, which generate intermediate interpretations that exist outside of the created artifact (e.g. images, margin summaries), intermediate text consists of suggestions that are part of the creative artifact itself but do not remain as-is in the final outcome.
The concept of intermediate text was inspired by the initial writing stages where writers jot down ideas before they iteratively revise their writing. As the process of writing not only changes the textual artifact but also writing goals, the final writing may be drastically different than the imagined outcome that was desired in the beginning.

\subsection{System Overview}
\label{sec:system_overview}

\textit{Ai.llude} is a Flask-based web application (Figure \ref{fig:interface}). It contains a customized Quill.js text editor \cite{quill}, where writers can compose stories and request AI suggestions, which are continuations of the text in the editor. Interactions with the editor and with the AI suggestions interface are comprehensively logged; complete writing sessions can be reconstructed.

To make the AI user-interface more streamlined for text editing, the editor displays a drop-down menu at the caret location when the tab key is pressed, inspired by CoAuthor \cite{leecoauthor22}.
The drop-down menu displays two possible AI suggestions: \textbf{Suggestion A} and \textbf{Suggestion B}, corresponding to \textit{fluent continuation} and \textit{intermediate suggestion}. These suggestion types are named generically in the interface to prevent nomenclature bias. Figure \ref{fig:interface} shows the drop-down menu.

All suggestions are requested and received using the OpenAI Chat completions API and generated with GPT-3.5 Turbo. Each request contains the story from the beginning to the current caret location. The returned suggestion is inserted after the caret in order to let AI suggestions be used anywhere in the story. If the provided story text is too short (less than 100 characters), the AI suggestions are disabled to reduce the likelihood of an irrelevant suggestion. Generated text is inserted in purple to visually distinguish human and AI text.  When the user types, the user-input characters are presented in black. 

\textbf{Suggestion A} provides AI generated text that is designed to be a \textit{fluent continuation}.  The text up to the caret is provided to the model for completion; no additional system or user prompts are provided, and no parameters are changed from the default (e.g. temperature = 1), so a fluent continuation mimics the style and tone of the story. As this suggestion provides the statistically optimal continuation of the text, we imagine that this type of suggestion has relatively \textbf{low interactivity}; i.e., less potential for revision and less room for critical engagement. We hypothesize that this type of AI generation does little to encourage rewriting. To prevent suggestions from overwhelming the writer and story, Suggestion A suggestions are terminated after a newline (signifying a new paragraph) or after 60 tokens are generated, whichever comes first.

\textbf{Suggestion B} inserts a sequence of unrelated ideas concatenated into a single string, designed not to be used as part of the written artifact as-is. It is designed to be a prototype of \textit{intermediate text}. 
Specifically, this suggestion queries the Chat completions API independently four times with prompts designed to return the following types of continuations:
\begin{enumerate}
    \item \textbf{Plot Beat}. Suggests a shift in narrative; e.g., new events, actions, emotional turns, realizations, etc. This is designed to suggest a new plot element that should be further expanded if used.
    \item \textbf{Setting Detail}. Adds a new detail about the setting, such as new sensory information or a new setting element; e.g., a flowering tree if the setting is a courtyard. This is designed to expand the writer's imagining of the scene and should be expanded if used.
    \item \textbf{Benign Tangent}. Suggests a benign circumstantial event occurring in the background; e.g., a squirrel scurrying across the grass in a fishing scene. This is designed to expand the writer's imagining of immersion and should be expanded if used.
    \item \textbf{Whimsical Storyteller}. Provides a continuation of the story in a light-hearted and joking tone. This is designed to be a stylized continuation of the story that should be rewritten into a more similar style to the story if used.
\end{enumerate}
Each individual query is instructed to cap itself at 15 words, with a 30 token hard limit, in order to approximate the length of Suggestion A. Table \ref{tab:suggestion_example} shows examples of both types of suggestions.

Suggestion B was developed through multiple iterations, with the initial goal of requiring editing to be grammatical. Early versions included removing punctuation and capitalization to create lengthy, ungrammatical run-on sentences inspired by stream of consciousness writing. However, pilot testing showed that this design impeded usability because it required too much effort to read and was too difficult to understand. 
To improve ease of use and relevance to creative writing, we decided to append multiple unrelated continuations together. This was designed to be easy to read, but still ill-formed to encourage rewriting. \textbf{Plot Beat} and \textbf{Benign Tangent} are designed to assist with the development of the story's direction, while \textbf{Setting Detail} and \textbf{Whimsical Storyteller} are designed to assist with crafting prose.

In general, we design intermediate suggestions to have \textbf{high interactivity} and to encourage \textbf{greater control} over creative decisions. First we imagine a selection process to take place: one component might be chosen and the rest might be discarded. Secondly, we imagine that due to the suggestions being underdeveloped or improperly stylized, any desired components will be further rewritten. We hypothesize that combining selection with rewriting will make the suggestion more interactive.

\begin{table*}[]
    \centering
    \begin{tabular}[t]{|L{0.35\textwidth}|L{0.35\textwidth}|}
        \hline
        \multicolumn{2}{|L{0.72\textwidth}|}{\textbf{User Text}} \\
         \multicolumn{2}{|L{0.72\textwidth}|}{Alice was beginning to get very tired of sitting by her sister on the bank, and of having nothing to do: once or twice she had peeped into the book her sister was reading, but it had no pictures or conversations in it, “and what is the use of a book,” thought Alice "without pictures or conversations?"} \\
        \hline
        \textcolor{blue}{\textbf{Fluent Continuation}} & \textcolor{orange}{\textbf{Intermediate Suggestion}}\\
        So, she decided to explore the riverbank. As she wandered, she stumbled upon a peculiar looking rabbit hole hidden behind some bushes. Curiosity piqued, Alice peered into the dark abyss, wondering where it might lead. With a deep breath, she made a bold decision to follow the white rabbit, diving... & Suddenly, a peculiar-looking creature appeared with a book that could bring pictures to life... Just as Alice was about to resign herself to the monotony... Suddenly, a squirrel scurried up to Alice and sat on her lap... The bank where Alice and her sister sat was lush with wildflowers of every color imaginable... \\
\hline
    \end{tabular}
    \caption{
        Two types of text completions can be requested. Top: user text. Bottom left: a potential fluent continuation. Bottom right: a potential intermediate suggestion. Figure user text: Alice in Wonderland (Lewis Carroll).
    }
    \label{tab:suggestion_example}
\end{table*}
\section{Methods}

To evaluate the effects of AI writing on rewriting, we asked creative writers to each write a short story using \textit{ai.llude}. 
We received IRB approval from our university in order to evaluate the system on human participants. Study risks were stated in the consent form: (1) privacy concerns of using an online system and (2) the potential risk to receive harmful AI-generated content. To help mitigate (1), participants were informed that they can prevent sending editor interaction data to the server by refreshing the editor without explicitly saving, and an enterprise OpenAI account was used to have complete ownership of all data. To help mitigate (2), a language filter was used to prevent harmful language returned by the Chat completions API from being displayed in the text editor. If a completion included a blacklisted word, it would not be shown.

\subsection{Participant Recruitment}
People who self-identified as creative writers and consented to the study were recruited from university mailing lists, fliers posted in public locations (e.g. a local book store), and by word-of-mouth. We screened study applicants using their reported creative writing history: we recruited participants who had previously engaged in creative writing for personal expression or enjoyment. Study applicants who did not previously engage in creative writing or only wrote personal essays (e.g. college application essays) were not included. Thirty-two participants were recruited for the study. Five participants did not write for the full session, or only pressed the AI suggestion button and did not write any text; these sessions were excluded from the data analysis. This left 27 participants who completed the entire study.

Of these, 11 identified as male, 14 identified as female, 1 self-described, and 1 did not disclose. The average age was 22.7 years (min: 18, max: 32). Nine participants self-identified as a beginner writer; 14 self-identified as an intermediate writer, and 4 self-identified as an expert writer. The three most commonly self-reported types of writing were short fiction (17), journals (12), and poetry (11). 
Fifteen participants had earned a bachelor's degree, and 12 were currently undergraduate students. While we did not restrict recruitment, most participants were college students or recently graduated from college, with an average of 6.3 years of writing experience.
Due to the limited age range of the sample set, and geographic similarity, our results are likely influenced by demographic characteristics.

In the pre-survey, participants mostly felt that AI text generators were \textit{good at generating fluent text} (agree: 21, neutral: 5, disagree: 1) and at \textit{providing useful responses} (agree: 19, neutral: 8), but participants were divided on whether AI text generators were \textit{good at providing new ideas} (agree: 7, neutral: 13, disagree: 7).
Study participants were compensated \$10 USD.

\begin{table*}
\begin{tabularx}{0.75\textwidth}{| @{\hspace{2em}}p{0.41\textwidth}@{\hspace{2em}} | p{0.3\textwidth}}
    \cline{1-1}
    \\
        \qquad "I mean, I think that maybe we should just take a break and see what it might mean. \textcolor{blue}{Maybe there's something we're missing, you know?" Theo suggests, his voice tinged with uncertainty.}   
        "Maybe we're both holding each other back from something." 
        &  \textcolor{blue}{\textit{ \newline \newline Fluent continuation.}} \\
        \qquad \textcolor{orange}{ Lucy's heart sinks as Theo's words hang in the air. A break? It feels like the ground beneath her has given way.} 
        & \textcolor{orange}{\textit{Intermediate suggestion, edited to keep only a portion of the suggestion.}} \\
        She has to \textcolor{blue}{look out the window} so that he doesn't see the tears accumulating in her eyes. \textcolor{blue}{The city seems to be in constant motion} while her world has just stopped.  
        &  \textcolor{blue}{\textit{Fluent continuation, rewritten to incorporate multiple segments of user text.}} \\
    \\
    \cline{1-1}
\end{tabularx}
\caption{An excerpt of participant W18's story, showing how they integrated both types of AI suggestions. User text is presented in black. Text generated by Suggestion A is in blue, and Suggestion B in orange. The portion of the intermediate suggestion kept in the story was the ``whimsical'' continuation. Note that in the \textit{ai.llude} interface, all AI text is presented in a single color.}
\label{tab:annotated-example-text}
\Description{Table 2 shows an example excerpt of user writing. User text is displayed in black. Fluent continuation text is displayed in blue. Intermediate suggestion text is displayed in orange. It contains some user text, followed by fluent continuation text, followed by user text, followed by intermediate suggestion text, finally followed by a mixture of user text and fluent continuation text.}
\end{table*}

\subsection{Experimental Design}
\subsubsection{Writing Environment}
\label{writing-environment-experiment}
To create a realistic and engaging writing environment, writers were asked to imagine a fictitious scenario of writing a short story to be submitted to a literary column. Short story writing is a common task for studying writing assistants \cite{calderwood20hownovelists, leecoauthor22, carrera22watch, clarkslogansstories18, roemmele_creative_2015, roemmele_automated_2018, singhstolenelephant22}; it is a task familiar to participants and relatively constrained. Participants were provided intentionally vague descriptions of both fluent and intermediate suggestions to encourage experimentation and completed a short tutorial session. In the tutorial, participants were encouraged to try both suggestion types on a provided story excerpt to acclimate to the writing environment and see examples of potential AI output. 
To help provide writing inspiration, three optional writing prompts were provided. Writers were instructed to write for at least 50 minutes. The AI suggestion menu was disabled for the first 15 minutes of writing, or until 150 words were written, to encourage writers to develop initial ideas and story-writing plans without AI influence.  To reduce order effects, the tab menu presented Suggestion A on top to half the participants, and Suggestion B on top to the other half. Writing prompts and user instructions are provided in the supplemental materials.

\subsubsection{Writing Evaluation Survey}
\label{writing-evaluation-survey}
At the end of the writing session, participants filled out an exit survey in which they reflected on their experiences using the AI suggestions. To evaluate the effects of fluent and intermediate suggestions, the exit survey probed overall feelings of satisfaction (e.g. "are you proud of your story") and self-concept ("e.g. does your writing sound like you") \cite{vandynePsychOwnership04, Belk88}. 
By giving the survey at the end of the writing session, these questions were designed to gauge attitudes towards the overall writing outcome and to capture the overall process of using and integrating AI assistance.

Participants were also asked to reflect on each suggestion type by considering self-concept ("The suggestion sounded like something I could have written") and responsibility ("I felt the need to rewrite what the suggestions gave me") \cite{vandynePsychOwnership04, Belk88}. This subset of questions was designed to assess whether suggestions were aligned with the writer's personal intentions and writing goals. Finally, participants were asked to reflect on how they used, evaluated, and integrated both type of suggestions as open-ended responses. The survey was informed by prior literature on psychological ownership and control \cite{vandynePsychOwnership04, pickfordPsychOwnership16}, which we chose as a way to assess self-expression. The full list of survey questions can be found in the supplemental materials.

\subsubsection{Rewrite Metrics}
We define rewriting as a series of actions that changes an original span of text to the version of that text in the final document. 
In this work, we are interested in the changes that people make when they rewrite AI suggestions. 
Therefore, we investigate rewriting by looking at the differences between the original AI suggestion and the final rewritten version.
Here, we describe both how we operationalize initial and final spans of text, and the metrics used to characterize the differences. 
We define the ``original suggestion'' as the span of text initially appended by the AI to the editor.  
We define
the final ``rewritten version'' as the span of text in the final draft that begins at the first remaining AI character from the original suggestion, and ends at the final remaining AI character, i.e. the suggestion after new text insertions and deletions. See Table \ref{tab:rewritten-example} for an example.
To evaluate the extent of rewriting, we measure the difference between the original AI suggestion and its final rewritten version using three metrics:
\begin{enumerate}
    \item \textbf{Words Remaining.} This is the percentage of AI text that was kept from the original suggestion in the final rewritten version; specifically, the number of AI words in the final rewritten version divided by the number of AI words in the original suggestion.
    \item \textbf{Sentence Embedding Distance.} This measures semantic similarity between the original suggestion and the rewritten version using cosine similarity of the two computed vectors calculated by \cite{reimers-2019-sentence-bert}. 
    \item \textbf{Number of Edits}. This counts the number of \texttt{text-insert} and \texttt{text-delete} operations occurring within each suggestion. The number of user edits is a measure of the effort it took to rewrite the original suggestion. Counting text operations is important for measuring rewriting since writers might make repeated changes that might not be captured by comparing the original and final states, such as typing and backspacing or undoing and redoing. 

\end{enumerate}

We calculate these metrics for all AI suggestions, whether A or B, and compare the results for these metrics between suggestion types.

\subsubsection{Qualitative Analysis}
To contextualize the quantitative results with how and why participants rewrote AI suggestions, we conducted a 
reflexive thematic analysis \cite{braun2006TA} on the open-ended responses of the exit survey (Section \ref{writing-evaluation-survey}). 
Because rewriting AI text is an underexplored area, inductive coding allows us to independently build up our understanding of rewriting behaviors from the data. 
Both authors coded a subset of survey responses, then discussed their codes to share interpretations and reach consensus. Then the first author open-coded the remaining responses and grouped the codes into potential themes. That author then checked each code and theme again to refine groupings and merge similar themes. The themes were discussed and refined again by both authors.

\begin{figure*}
    \centering
    \includegraphics[width=0.95\textwidth]{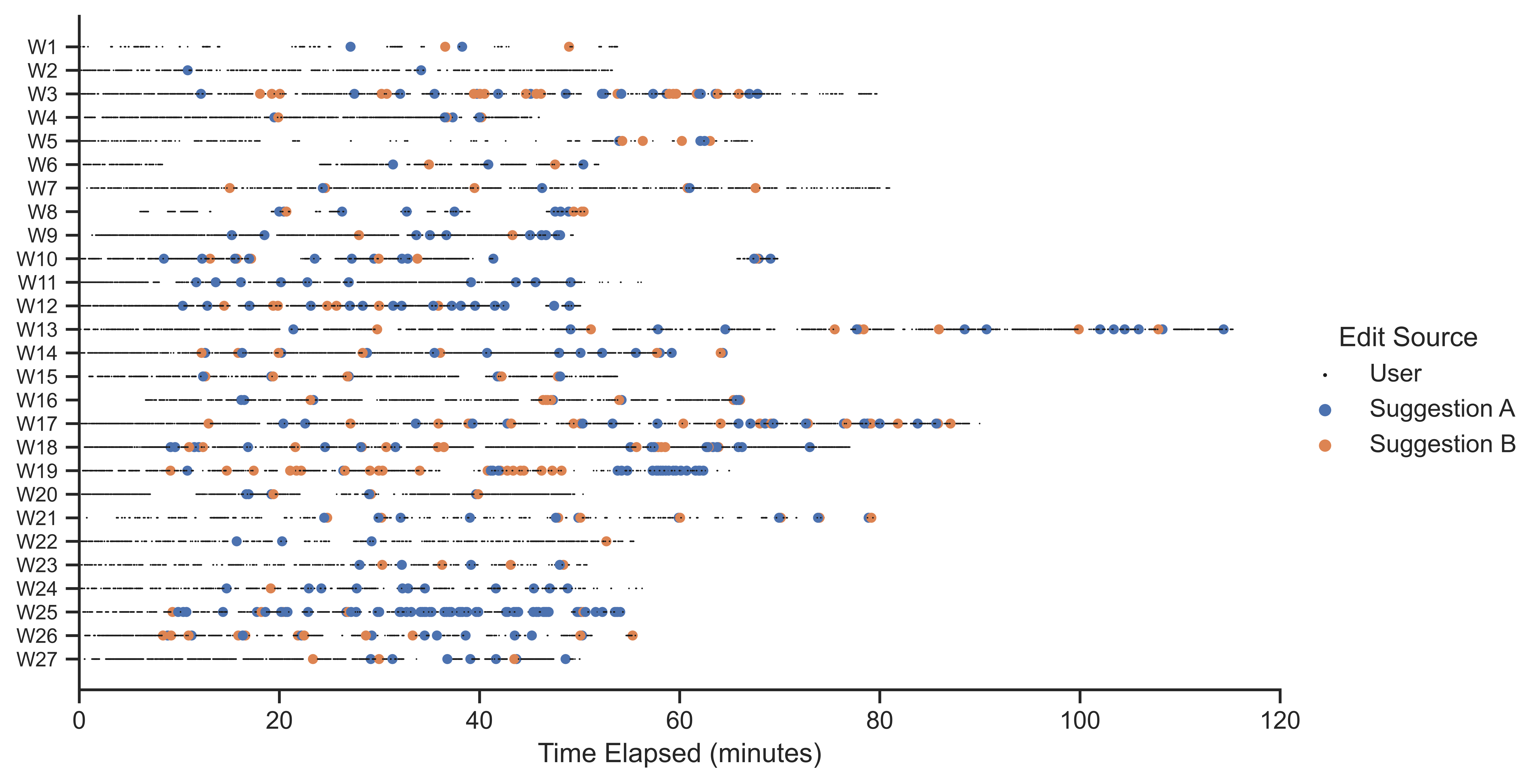} 
    \caption{Writing Flows.  Dots indicate activity in the text editor, either of the user typing or the user requesting an AI suggestion. The suggestions were disabled for the first 15 minutes of the session, or until 150 words had been written.} 
    \label{fig:writing_history}
    \Description{Figure 4 shows the writing flows of each participant. There are 27 rows, each representing a participant. Black dots on the row represent user edits; blue dots represent usages of Suggestion A; orange dots represent usages of Suggestion B. There are 27 rows, each representing a participant. The x-axis shows time elapsed in minutes and the y-axis shows participant ID, from W1, W2, …, W27. }
\end{figure*}

\section{Results}
We analyze the data collected from our study in three ways.  First, we characterize overall writing and rewriting behaviors of participants when using \textit{ai.llude}, finding two predominent themes of AI text rewriting: \textit{stylistic consistency} (Section \ref{stylisticconsistency}), and \textit{selection process} (Section \ref{selectionprocess}).  Then, we compare writing and rewriting behaviors between fluent continuations and intermediate suggestions, and show that the design of intermediate suggestions significantly increases rewriting (Sections \ref{BRW} \& \ref{sec:B-motivates-rewriting}).
We address the relationship between suggestion design and writing process, finding that fluent continuations make writing easier and intermediate suggestions support reflective thinking (Section \ref{sec:ease-vs-reflection}). Finally, we report how suggestions were adapted in sometimes unintended ways, an illustration of creative misuse (Section \ref{sec:creative-misuse}).

\subsection{Characterizing Writing Process and Rewriting Behaviors}
\label{sec:characterizing}
Stories written with \textit{ai.llude} were on average 1,205 words long, equivalent to approximately 2 pages of 12pt single-spaced text (min: 218 words, max: 2,969 words). Stories contained on average 370 AI generated words, i.e. 31\% of total text (min: 0 words, max: 2,197 words). The average writing session lasted for 62.7 minutes (min: 46.0 minutes, max: 115.3 minutes). Table \ref{tab:annotated-example-text} shows an excerpt of a participant-written story, demonstrating how a writer might integrate suggestions into their story and how the final text alternates between AI-generated and user-written content.

Figure \ref{fig:writing_history} depicts timelines of each writing session, indicating when the user triggers AI suggestions and when the user types. Some writers traded off with AI suggestions and requested them sparingly, with lengthy periods of their own writing in-between each suggestion (W2, W7, W16). Others used suggestions more frequently, sometimes in bursts, for example deleting unhelpful suggestions until landing on a useful one (W3, W19, W25). Writers varied in their usages of both suggestion types: one writer used Suggestion B once, then stuck with Suggestion A for the rest of the writing session (W24); one used Suggestion B more often (W19). Others tended to switch between both types equally.

After requesting AI suggestions, writers take one of three actions: 1) they delete the entire suggestion, 
2) they leave the entire suggestion as-is, or 
3) they rewrite part or all of the suggestion, including deleting portions of the suggestion. 
Suggestions with one segment contain no internal user segments; e.g., text left as-is (i.e. one AI segment and no user segments). Writers would also oftentimes delete large portions of generated text and use a small but contiguous piece, which is also one AI segment and no user segments.
Fewer segments are more common; 92\% of suggestions that are kept in the text include five or fewer segments.  
Table \ref{tab:rewritten-example} shows an initial AI continuation provided by a fluent continuation (surrounding user text provided for context), which was then rewritten by the user.
This rewriting resulted in nine alternating segments: a large insertion describing the narrator's backstory, followed by small grammatical edits to set the suggested setting description to past tense.

\begin{table*}
    \centering
    \begin{tabularx}{\textwidth}{| @{\hspace{2em}}p{0.35\textwidth}@{\hspace{2em}} | @{\hspace{2em}}X@{\hspace{2em}} |}
    \hline
    \textbf{Initial Suggestion}\newline & \textbf{Rewritten Suggestion}\newline\\
            
            ... because that was some much needed confidence boost for the nerve-wracked \textcolor{blue}{me. I take a deep breath and straighten my tie, making sure every detail is perfect. The Rabbit Diner is a quaint little place, known for its cozy atmosphere and delicious comfort food. As I step inside, the warm aroma of coffee and pancakes envelops me, instantly calming my nerves} and bringing my attention to the more important matter at hand, my blind date.   
        & 
            
            ... because that was some much needed confidence boost for the nerve-wracked \textcolor{blue}{me. I take a deep breath and straighten my tie, making sure every detail is perfect.} I almost get a heart attack with a almost perfectly timed beep from my pager. Yeah you heard (or read) that right. A beep from my PAGER. Call me a old soul or whatever but nothing screams security like good ol'fashioned technology. I dont have space in my head for tiktok or trends. Throw any language or subject at me, and I can bore you to death with knowledge about it. You want me to dance like a toy at the whims of the current trends, both literally and figuratively? Count me out. "Red rabbit at yellow gates at full moon ", the pager states. That immediately distracts me from my nerves. Apparently, my life-long nemesis(long story short, we hate each other and work for opposite sides) is a lady with red hair and she has a meeting with some important clients at the gates of the Pakistan museum at noon. How did they even discover all of this information when I was never able to even catch a glimpse at their face for all these years?
        
            \qquad \textcolor{blue}{The Rabbit Diner} was \textcolor{blue}{a quaint little place, known for its cozy atmosphere and delicious comfort food} and a\textcolor{blue}{s I step}ped \textcolor{blue}{inside, the warm aroma of coffee and pancakes envelops me, instantly calming my nerves} and bringing my attention to the more important matter at hand, my blind date.    
         \newline
         \\
         \hline
    \end{tabularx}
    \caption{Two versions of an excerpt of participant W9's story.  Left: An AI suggestion inserted into the user text.  Right: The same excerpt after the user rewrote the suggestion, including a large insertion of character backstory, and small grammatical edits. Text from the original suggestion is presented in blue. Surrounding user-written text provided for context.}
    \label{tab:rewritten-example}
    \Description{Table 3 has two cells. The cell on the left contains the initial suggestion text; AI text is purple and user text is black. The cell on the right contains the rewritten suggestion text; AI text is blue and user text is black. The rewritten text contains a large segment of new user text in between the AI text, followed by small user edits.}
\end{table*}

\subsubsection{Stylistic consistency}
\label{stylisticconsistency}
AI suggestions were helpful for introducing novelty, inducing effective surprise in the writer.
To integrate suggestions into their stories, writers would keep the core concept and rewrite the suggestion to make it stylistically fit with the writer's intentions. This type of rewriting included adapting the text to the writer's own voice
, changing the style of text to align with the writing subject
, and ensuring consistency with surrounding style:
\squote{W4} {I merely changed the suggestion to make it sound a bit more like what I would write.}
\squote{W1} {If I liked the suggestion I would make it stylistically fit in with the prompt.}
\squote{W10} {I usually reworked any of the suggestions by modifying the words used to fit the formality of the passage or by reworking its flow so that it sounded better.}

Conversely, some writers also used AI suggestions to seek wording and style, discarding the idea or direction contained in the suggestion:
\squote{W1} {If I didn't like the ideas I would see if I like the style of writing could fit.}
\squote{W18} {For instance, if I needed the name of something, I would use Suggestion A and regenerate it until it produced a name I liked.}

Generally, the value of a suggestion depended on its \textbf{novelty} (whether it contained new ideas, directions, alternatives, or interesting plot elements) and its \textbf{stylistic coherence} (whether it aligned with writing style, tone, voice, flow, and creative intentions). Suggestions did not need to contain all of these elements to remain in the story or be useful to the writer. Ones that contained novel ideas, but with undesired style, were integrated and rewritten for stylistic consistency; others with relevant style, but uninteresting ideas or direction, were used to assist with wording. These insights build off of Singh et al.'s findings that in some cases, writers were not hindered by less semantically relevant suggestions or less linguistically coherent sentences \cite{singhstolenelephant22}.

\subsubsection{Selection process}
\label{selectionprocess}
AI suggestions, particularly the intermediate suggestions provided by Suggestion B, tended to contain elements that were not all useful for the writer. This would result in a selection process: writers used suggestions to consider smaller phrases and alternatives, and chose the pieces that they wanted and discarded the rest. The selection process could, but did not necessarily, co-occur with integrating style; the selection process also helped writers consider alternatives and potential directions for their story:
\squote{W4} {I picked and chose those which could help me better my story.}
\squote{W7} {I did use fragments of some suggestions (I think mostly Suggestion A) to fill in phrases I needed but couldn't come up with.}
\squote{W19} {"B" often had one or two usable (given the context) possibilities, but went into a lot of irrelevant and generic detail.}

The selection process was not limited to immediate integration of the suggestions. Some writers also selected pieces of suggestions for \textit{distant use}:
\squote{W6} {Suggestion B was usually a bit more messy and incoherent, so I would split up Suggestion B and maybe use it different places.}
\squote{W10} {If suggestion B had a part that I liked but didn't fit then I would make a note of it to use later in the story.}

Similar to rewriting for stylistic consistency, the value of a suggestion depended on novelty and stylistic coherence, and did not need both to be fragmented and selected (choosing fragments) by the writer. The selection process enabled writers to assess and integrate suggestions \textbf{recursively}; if an entire suggestion was unusable as-is, there could still be value in integrating a smaller, but more useful fragment, while deleting the rest. 
Thus, the selection process surfaced as a subprocess of integrating AI suggestions.

\subsection{Assessing Usage of Fluent Continuations vs Intermediate Suggestions}
\label{sec:AvsB}

\subsubsection{Intermediate suggestions increase rewriting}
\label{BRW}

\begin{figure*}
    \centering\
    \subfloat[\centering Remaining words. Higher values correspond to more AI words kept.]{{\includegraphics[width=0.45\textwidth]{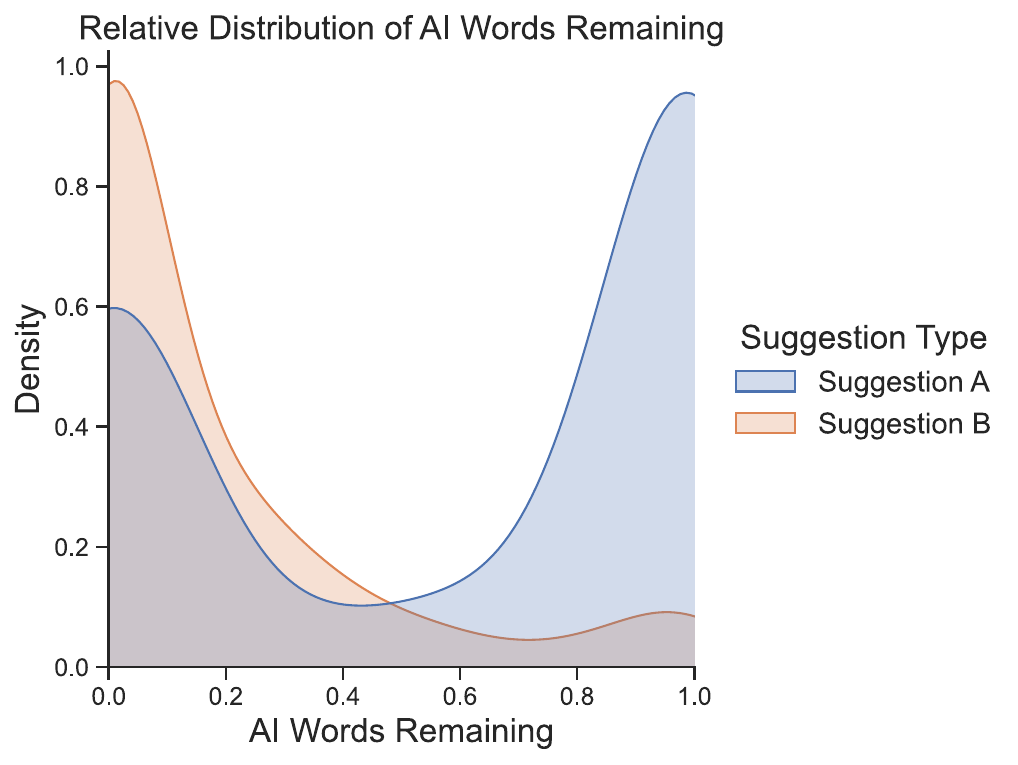} }}
    \quad
    \subfloat[\centering Sentence embedding similarity. Higher values correspond to greater similarity.]{{\includegraphics[width=0.45\textwidth]{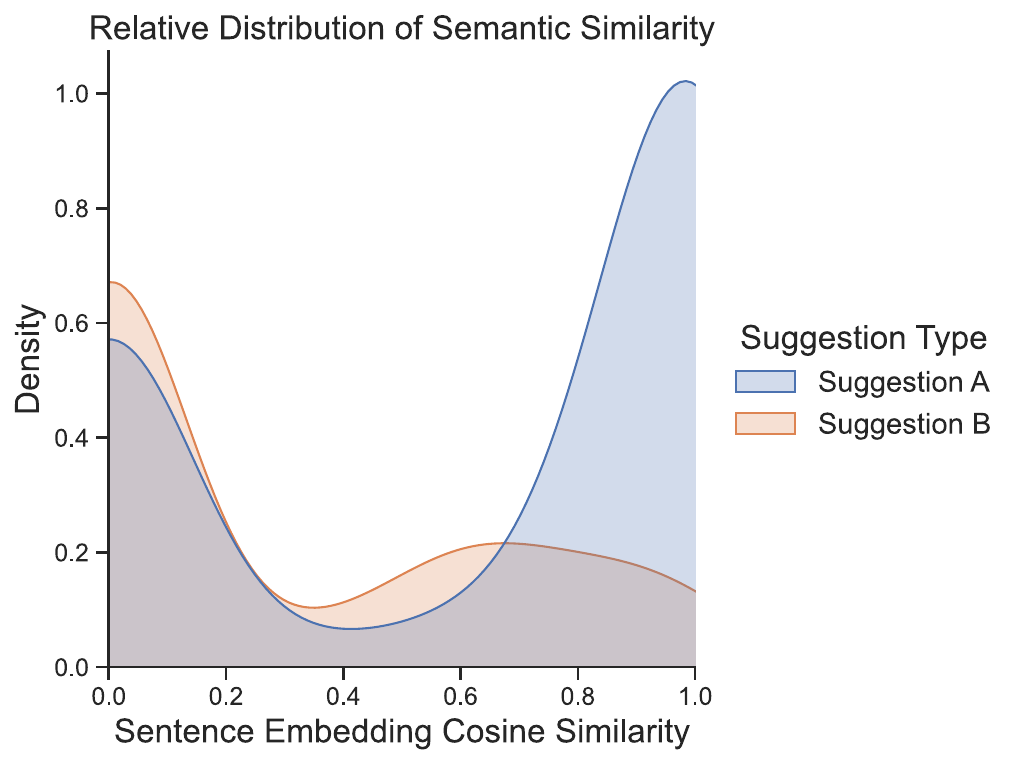} }}
    \quad
    
    \caption{    
    Suggestion A (fluent continuation) vs. Suggestion B (intermediate suggestion) average rewrites. Left: Rewrites of Suggestion A tend to have more AI text remaining compared to rewrites of Suggestion B. Right: Rewrites of Suggestion A tend to be more similar to their initially provided forms compared to rewrites of Suggestion B. }
    \label{fig:A-B_Rewrite}
    \Description{Figure 5 shows two subplots. The left subplot compares the remaining words in Suggestion A suggestions to the remaining words in Suggestion B suggestions. It is a kernel density plot, with the x-axis representing words remaining from 0.0 to 1.0 and y-axis representing density (relative frequency). Most of suggestion A is kept; most of suggestion B is deleted. The right shows a similar plot as the right, but the x-axis displays sentence embedding cosine similarity. Rewrites of Suggestion A tend to be very similar to the original text (higher density closer to 1), while rewrites of Suggestion B tend not to be similar to the original text. }
\end{figure*}

Writers tended to keep Suggestion A in its entirety, rewrite it to some extent, or reject it, whereas very few Suggestion B suggestions were kept in their entirety.
Figures \ref{fig:A-B_Rewrite}a and \ref{fig:A-B_Rewrite}b compare original and rewritten suggestions according to remaining words (0: no words kept; 1: all words kept) and sentence embedding similarity (0: not semantically similar; 1: very semantically similar). 
\textbf{Writers kept significantly less of intermediate suggestions than fluent suggestions, both on page and semantically}: fluent suggestions had an average of 60\% remaining text and intermediate suggestions had an average of 16\% ($p=0.00, t=11.51$). Using sentence embedding similarity, fluent continuation had an average of 0.63 and intermediate suggestions had an average of 0.28 ($p=0.00, t=8.70$). 

The average text operation count (number of \texttt{text-insert} and \texttt{text-delete} events) between the provided suggestion and the final text was significantly different for fluent continuations and intermediate suggestions as well (\textit{fluent continuation}: 444, \textit{intermediate suggestion}: 795, $p=0.001, t=-3.09$). \textbf{Writers made nearly double the number of edits for intermediate suggestions compared to fluent continuations.}
Designing AI suggestions to provide intermediate solutions significantly changes how writers engage with the suggested text and results in a more user-written and effortful final outcome. 

\begin{figure*}%
    \centering
    {\includegraphics[width=0.95\textwidth]{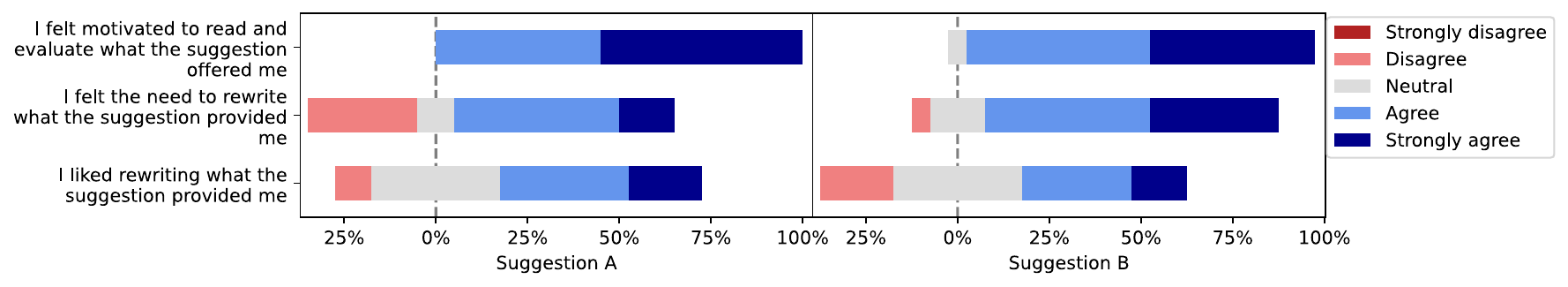} }
    \caption{Writers were similarly motivated to read and evaluate both Suggestion A and B text, but more writers felt the need to rewrite Suggestion B.}
    \label{fig:responsibility}
    \Description{Figure 6 shows the distribution of responses to 3 likert questions related to perceived responsibility. The plot shows that most participants felt motivated to read and evaluate what the suggestion offered; more participants felt the need to rewrite Suggestion B compared to Suggestion A; and most participants generally enjoyed rewriting A and B.}
\end{figure*}

\subsubsection{Intermediate suggestions motivate rewriting}
\label{sec:B-motivates-rewriting}
Figure \ref{fig:responsibility} presents survey responses related to responsibility \cite{vandynePsychOwnership04} for fluent continuations and intermediate suggestions. The perception of Suggestion B as lower quality was reflected by the greater perceived need to rewrite it. While there was no significant difference regarding motivation to read and evaluate what the suggestion offered (writers would read and think about the suggestion), significantly more writers reported that they felt the need to rewrite intermediate suggestions ($p=0.025, U=268.5$). To test for significance on ordinal data without assuming normality, the Mann-Whitney \textit{U} test was used for Likert scale responses.

\subsection{Mediating Ease vs Reflection}
\label{sec:ease-vs-reflection}
In Section \ref{BRW}, we empirically observed significantly more rewriting when using intermediate suggestions. Here, we analyze the perceived differences between both types of suggestions and find that fluent continuations are easier to integrate and useful for getting unstuck, while intermediate suggestions are useful for redirecting writing and supporting reflection.

\subsubsection{Fluent suggestions mediate ease}

Writers liked Suggestion A for requiring less effort for evaluating and integrating its suggestions. The fluent continuation was designed to be coherent with the rest of the story and able to be left as-is, and unlike Suggestion B, presented only a single possible continuation. Due to these reasons, writers reported \textbf{needing to make fewer changes} to integrate fluent suggestions:

\squote{W4} {I merely changed the suggestion to make it sound  a bit more like what I would write.}
\squote{W12} {Suggestion A created a single new idea and wrote a cohesive and integrated paragraph about it. }
\squote{W13} {I think I overall liked Suggestion A better, ... I could more easily rework to fit in the paragraphs.}

Conversely, Suggestion B required more effort to use, as it was ungrammatical and presented four possible continuations. Oftentimes, writers reported needing to spend more effort rewriting intermediate suggestions compared to rewriting fluent continuations.
\squote{W6} {Suggestion B was usually a bit more messy and incoherent.}
\squote{W10} {I would do the same for suggestion B but there was a bit more rewriting and expanding on the ideas necessary from my side.}

Fluent continuations also helped to mediate ease by helping writers get unstuck. They alleviated struggle by directly inserting text that satisfied the writer's needs, moving them along to the next writing task. These writing difficulties tended to relate with \textbf{wording}:
\squote{W14} {I ... used suggestion A when I slowed down or got stuck.}
\squote{W18} {The suggestions made the writing process faster, especially when using Suggestion[s] A, as those provided quick details or vocabulary that added what I wanted to the story.}
\squote{W22} {A helped with adding details and just kind of moving the story along when I was stuck, so I think I ended up just leaving it as is.}

Generally, using fluent continuations to mediate ease did not result in as much rewriting. Since writers used fluent continuations to streamline their writing and reduce struggle, there was less motivation to rewrite them beyond what was needed for a stylistically consistent integration. 

\subsubsection{Intermediate suggestions support reflection}
On the other hand, writers liked Suggestion B for helping with process and creative decision making. While fluent suggestions provided only a single continuation, intermediate suggestions provided potential directions for writers to consider; specifically, by \textbf{comparing alternatives}:
\squote{W3} {Suggestion A helped me with the descriptions and details, while suggestion B helped me mostly with thinking about the different actions and how to consider different alternatives in making up this story}
\squote{W11} {Better for generating ideas than A but not well written.}
\squote{W24} {Suggestion B was useful for trying to compare different paths for the plot of my story.}

By providing less coherent and shorter continuations, Suggestion B was used as a platform for evaluating personal writing goals. Suggestion B also required more rewriting to adapt, which allowed writers to integrate more of their own writing:
\squote{W1} {Suggestion B I thought was more interesting and useful because it gave me another platform to speak on. }
\squote{W10} {with Suggestion B I often had to use that as a launch pad and rewrite it completely in my own words using the ideas presented.}
\squote{W23} {Between Suggestion A and Suggestion B, I was more likely to keep Suggestion B since it had more, shorter, options which made it easier for me to evaluate new ideas but still integrate more of my own writing into the story.}

In general, writers used AI suggestions to assist with the difficult parts of writing, which differed between writers. This resulted in some writers developing preferences for one type of suggestion over the other. The suggestions that were the most useful engaged with the writer's intentions and process; these suggestions either helped the writer \textbf{continue writing their story} by providing a satisfactory continuation or \textbf{support thinking about creative decisions} by providing possibilities for the writer to consider.  

\subsection{Creative Misuse}
\label{sec:creative-misuse}
Rewriting was designed to be a proxy for control by encouraging writers to rewrite the AI output (thereby supporting their control over their expression) but as we observed, it was not the only way writers sought control. Some writers used the tools in unintended ways, and reappropriated the tools to further their own storytelling goals:

\squote {W13}  {I do know I slightly 'misused' the A.I tool early on, thinking that the program simply loved ellipsis.}
\squote{W13} {Suggestion B was strangely repetitive, which worked for and helped me develop the 'character' of the A.I. in my larger story.}

\begin{table*}
\begin{tabularx}{0.9\textwidth}{| @{\hspace{2em}}p{0.5\textwidth}@{\hspace{2em}} | p{0.3\textwidth}}
    \cline{1-1}
    \\
        \qquad \textcolor{orange}{As I settle into the hanging basket chair on my balcony, a mysterious package arrives at my doorstep....} \textcolor{blue}{Text generation failed. Try again!} Well.
        &  \textcolor{orange}{\textit{Intermediate suggestion.}} \newline \textcolor{blue}{\textit{Fluent continuation.}} \\

        \qquad   Let me be fair. I have only just now encountered this particular brand of A.I, and I obviously am not a skilled practitioner with this tool. 
    \\
    \cline{1-1}
\end{tabularx}
\caption{An example of intermediate suggestion design and a system error message being reappropriated for story development. User text is presented in black; text generated by Suggestion A is in blue, and Suggestion B in orange. Note that in the \textit{ai.llude} interface, all AI text is presented in a single color.}
\label{tab:suggestion-reappropriation}
\Description{This table shows an instance of the suggestions (an intermediate suggestion and a system error message) being reappropriated by the writer for storytelling purposes. It begins with an intermediate suggestion (in orange), then a fluent continuation (in blue), then user text (in black).}
\end{table*}

For instance, W13 wrote a story involving an AI character and used the ellipses in intermediate suggestions as a way to establish that character's voice, despite the ellipses being placed to separate each continuation for ease of reading. W13 even integrated a system error message into their writing, even though the message was used to indicate that text generation failed due to a failed AI request. Table \ref{tab:suggestion-reappropriation} shows how W13 integrated these elements into their story.

Although the suggestions were directly inserted into the text (helping the writer engage with their task environment \cite{flowerhayes81}), some writers used intermediate suggestions strictly for writing inspiration, deleting the entire suggestion after it appeared. These authors used the suggestions as \textit{resources for interpretation}, discussed in Section \ref{sec:writing-support-tools}:
\squote{W14} {I tried out suggestion B and it did help me think about what I wanted, but I never incorporated suggestion B.}
\squote{W22} {Even if I ended up erasing the suggestion, I also just liked seeing what it would come up with, even if it wasn't headed in the direction I was intending, since it gave good sensory details and ideas.}
Thus generated suggestions were used differently depending on the needs of the writer. Writers have diverse needs \cite{Ippolito22}: some expressed wanting longer continuations (up to several pages) to help think about story direction; some wanted only a single word or phrase for word-finding and prose. Since creativity is open-ended \cite{amabilesocialpsy83} and due to the myriad ways a writer might realize their writing goals, system designers should be aware of how their tool integrates into creative process and design tools that work flexibly outside of their intended use cases.
\section{Discussion}

\subsection{AI Writing Tools and the Role of Rewriting} 
Rewriting is an opportunity to critically reflect on what was provided within the context of the story and to transform it so that the story accurately reflects what the writer wishes to express.
In our study, we found that style and story direction were important considerations when AI-provided text was being integrated.
Since intentions, processes, and goals are not surfaced to text generators, the generated text is not necessarily aligned with the writer's needs and interests. 
Past literature has shown that storywriters can be driven by the desire to maintain a sense of personal writing identity \cite{biermann22}. Because creative writing is so closely tied to self-identity and meaning-making \cite{Murray91}, we argue that how AI writing is rewritten and integrated in creative work is an imperative consideration when designing AI writing tools. 
Rewriting is a crucial step in integrating AI suggestions into the written piece, which should ultimately should express writing intent.

Writing is a reactive process between textual artifact and goals \cite{flowerhayes81}, and rewriting presents opportunities to discover and synthesize new ideas \cite{morley2007cambridge}. As AI text generators have the potential to influence personal opinions in a co-writing context \cite{jakesch2023cowriting}, the role of rewriting is not only integrating the suggestion text itself, but also helping the writer develop intention. 
Both Suggestion A and B provided new possibilities for the writer to explore. 
In our study, we found that writers would often use AI suggestions, particularly intermediate suggestions, as a platform for reflection and to make plans. Writers would sometimes use the suggestion to scaffold future writing or revise previously written text. These strategies, along with rewriting the suggestion text itself, were not mutually exclusive per suggestion.

Past literature has identified that writers, particularly those who value the emotional fulfillment of writing, desire control when co-creating with an AI writing assistant \cite{biermann22}. Control is a component of psychological ownership \cite{Belk88}, and we see rewriting as a potential way for writers to seek ownership while co-writing with AI.
In our user study, we observed weak correlations between rewriting and psychological ownership: the more AI text is in the story, the less perceived ownership the writer would feel. This was predicted by Lee et al. \cite{leecoauthor22}, in their comparisons of editing actions across the entire document with ownership. Our data aligns with Lee et al.'s finding, with a Pearson correlation coefficient of $r=0.2$ (Lee et al.: $r=0.3$). When evaluating only within AI suggestions, we found $r=0.3$. When comparing the number of editing operations (note that Lee et al. count \texttt{text-delete} and \texttt{cursor} events while we counted \texttt{text-insert} and \texttt{text-delete}) and ownership, we found $r=0.1$ (Lee et al.: $r=0.0$) \cite{leecoauthor22}.
However, the connection of user-writing and user-rewriting to psychological ownership remains unclear. Neither our study nor Lee et al.'s conclusively shows the existence or lack of relationship.

One direction for more detailed studies is improving metrics for assessing psychological ownership. 
Literature on psychological ownership suggests subdividing attitude and self-concept, which teases apart the impact of factors such as positive feelings and satisfaction (attitude) from the sense of the object as part of the self (self-concept) \cite{vandynePsychOwnership04}. We suggest that further research into the separate components of psychological ownership may be beneficial for understanding the effects of design choices of AI writing tools on the processes of co-writing with AI. Much of the current literature on ownership is centered on \textit{possession}, and not necessarily \textit{creation}, which is an important factor when considering creative pursuits.

\subsection{Imagining Future Intermediate Text} 
Suggestion B used a predetermined list of suggestion prompts (Section \ref{sec:system_overview}). However, such fixed lists will not be relevant for all creative writing tasks, environments, or writers; therefore, personalizing outputs might help with usability and relevance of intermediate text. Some writers used AI suggestions to help craft descriptive prose. We imagine that a potential implementation might offer a palette of literary devices and idioms that a writer can choose from and integrate into their writing.

The effectiveness of intermediate text design is contingent on two principles: \textbf{relevance to writing process} and \textbf{creative value}. Suggestion B was helpful for planning and scaffolding writing: some writers would request a suggestion and keep majority of it but write significant amounts of text in between. For these writers, Suggestion B was useful for the planning process, the act of forming an internal representation of the knowledge that would be used in writing \cite{flowerhayes81}. Suggestion B was also used for reflection and thinking about alternatives.
The design of intermediate text should motivate the reviewing process, the act of evaluating and revising \cite{flowerhayes81}. Early iterations of Suggestion B discouraged evaluation and revision because the lack of punctuation and capitalization detrimentally hindered ease of reading. Suggestion B was then revised to provide grammatical segments while keeping the entire continuation itself ungrammatical to encourage both evaluation and revision.

Creative value, which comes from the suggestion's ability to induce effective surprise \cite{bruner62conditions}, requires intermediate text to provide relevant, yet novel suggestions for the writer. Suggestion B offered four segments that were designed to be relevant to short fiction writing, including setting and plot development. Although the novelty of each segment itself will depend on the capabilities of the underlying language model, we argue that designing intermediate text to surprise and inspire the writer will help decrease the reliance on pure serendipity of text generation that may be otherwise distracting and potentially frustrating to read.

\subsection{Limitations and Future Work}
Since users perceived the current design of intermediate suggestions as lower quality and not as easy to use (leading to less motivation to request it), one future direction is to explore how to improve user perception while retaining the increase in rewriting behaviors. Using personalized prompts based on individual writing needs and style may encourage rewriting more often by providing more relevant suggestions. Future systems should be more aware of what type of suggestion is useful for the writer \cite{calderwood20hownovelists}. Future explorations of intermediate text prompts might provide user control over the design of the suggestions and adaptability based on the writing context.

\textit{Ai.llude} was designed to record complete writing sessions; live replays can be constructed from the log data. Watching writing replays had facilitated self-reflection by helping writers recognize writing habits that they were previously unaware of; many writers were surprised how much effort they put into finding the right expression by rewriting \cite{carrera22watch}. Studying how writers specifically view themselves rewriting AI suggestions may help writers reflect on how they integrate AI text and lead to a richer characterization and analysis of rewriting behaviors.

Participants in our study wrote for approximately an hour on a specific prompt; while this allowed us to collect a broad set of user sessions, this may not be a sufficiently authentic writing session to deeply probe feelings of ownership in the long term or on deeply meaningful personal projects. Writers who spend more time and effort on long-form writing may feel differently about their work and process, particularly as they become used to the eccentricities of AI text generators.

\section{Conclusion}

Intelligent writing support tools impact how we write in professional, personal, and creative contexts, and will likely continue to grow in frequency and capability. 
Since process is important to human flourishing through doing creative work as well as to the outputs of creative work, it is essential that we consider how the tools that we design will impact process.
One way to characterize this effect is creators' perceptions of control over their writing process and rewriting behaviors when working with AI assistance.  
In this study, we built a tool to investigate the rewriting of AI-generated text, since rewriting is a key stage of the creative writing process.
We find that we can increase rewriting by designing AI suggestions to produce \textit{intermediate text}, i.e. text with targeted imperfections that cannot be incorporated as-is into the final output. 
We also find that writers used fluent continuations and intermediate suggestions in different and diverse ways.
We hope that designing for targeted and imperfect text can be a generative frame for writing support tool designers to focus on process.
We envision a future where AI-enhanced tools for creativity are designed primarily to nourish the human creative work of self-examination, expression, and growth.

\bibliographystyle{ACM-Reference-Format}
\bibliography{reference}

\end{document}